# Development of ATLAS Liquid Argon Calorimeter Front-end Electronics for the HL-LHC

**Tiankuan Liu, on behalf of the ATLAS Liquid Argon Group**

*Southern Methodist University,*
*Dallas, TX 75275, USA*
*E-mail*: tliu@mail.smu.edu

ABSTRACT: The high-luminosity phase of the Large Hadron Collider will provide 5-7 times greater luminosities than assumed in the original detector design. An improved trigger system requires an upgrade of the readout electronics of the ATLAS Liquid Argon Calorimeter. Concepts for the future readout of the 182,500 calorimeter cells at 40-80 MHz and 16-bit dynamic range and the developments of radiation-tolerant, low-noise, low-power, and high-bandwidth front-end electronic components, including preamplifiers and shapers, 14-bit ADCs, and 10-Gb/s laser diode array drivers, are presented in this paper.

**Contents**

**1. Introduction**

The LHC high-luminosity upgrade [1-2] in 2024-2026 requires the associated detectors to operate at luminosities of up to $7 \times 10^{34}$ cm$^{-2}$s$^{-1}$. This corresponds to a signal pile-up from up to 200 events per proton bunch-crossing, with the goal of accumulating a total integrated luminosity of 3000 fb$^{-1}$. In order to be able to retain interesting physics events even at rather low transverse energy scales, increased trigger rates are foreseen for the ATLAS detector [3]. At the hardware selection stage, acceptance rates of 1 MHz are planned, combined with longer latencies up to 60 μs to read out the necessary data from all detector cells. Under these conditions, the current readout of the ATLAS Liquid Argon (LAr) Calorimeters [4-6] will not provide sufficient buffering and bandwidth capabilities. Furthermore, the expected total radiation doses [2] are beyond the tolerance range of the current front-end electronics [7]. For these reasons, a replacement of the LAr front-end and back-end readout system is foreseen for all 182,500 readout channels, with the exception of the cold pre-amplifier and summing devices of the hadronic LAr Calorimeter.

The block diagram of the current front-end readout electronics [4-5] is shown in Figure 1. A thick-film hybrid preamplifier (PA) amplifies the signal of each analog detector cell transmitted via a 25-Ω or 50-Ω cable. Three shapers with three different (low, middle and high) gains further amplify and filter the output signal of each PA in order to meet the dynamic range requirements. A switched-capacitor array (SCA) analog pipeline chip samples a shaped signal at the LHC bunch-cross frequency of 40 MHz and stores it in the analog form during the Level-1 trigger latency. An analog multiplexer (MUX) selects consecutive four or five samples per channel for the events accepted by the L1 trigger from the SCA with the optimal gain scale. A 12-bit analog-to-digital converter (ADC) digitizes the samples. The serializer multiplexes digital data of ADCs and the optical transmitter transmits optically out of the detector via a single 1.6 Gigabit per second (Gb/s) optical link.

A trigger-less readout scheme has been proposed for the HL-LHC upgrade [1-2]. In this scheme, the analog pipeline and the gain selection related to the Level-1 trigger are removed and all shaped analog signals are sampled, digitized, and transmitted from the detector to the back-end electronics. The block diagram of the front-end electronics of the proposed trigger-less readout scheme is shown in Figure 2. Instead of using three shapers with low, middle and high gains and a 12-bit ADC for each detector channel in the current readout electronics, two shapers with low and high gains and two 14-bit ADCs for each channel are proposed to cover the 16-bit dynamic range. The sampling rate may increase from 40 MHz to 80 MHz to improve the noise performance. The current back-end electronics will be replaced to read out the new front-end electronics. The transverse energy and bunch crossing time reconstruction of all calorimeter cells are performed and the full spatial granularity detector information is used to enhance further discrimination against background. This paper will focus on the development of the front-end readout electronics.

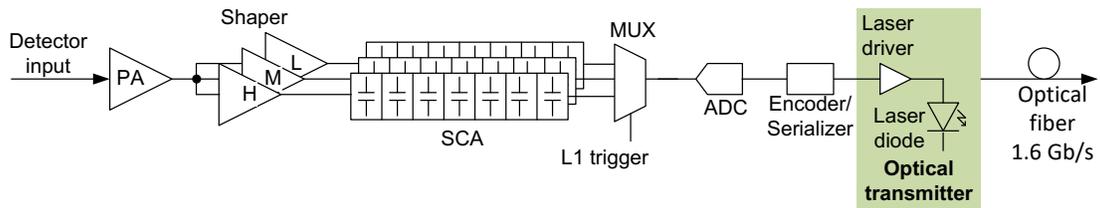

**Figure** 1. Block diagrams of the current front-end electronics

The goal of the front-end electronics upgrade is to maintain the same dynamic range (16 bits) and energy resolution (better than 0.25% at high energy) and to keep the same power consumption (about 630 mW per channel) as the current system. To meet this goal, a Front-End System-On-Chip (FESOC) approach is being pursued to reduce the system complexity and power consumption. In this approach, an ASIC integrating 8 channels of PAs and shapers (high and low gains on each channel), 16 channels of 14-bit ADCs, and a multiplexer/encoder/serializer unit is being targeted. The optimal scenario is to integrate all functional blocks on a single silicon die. If this goal cannot be achieved, a single package with more than one die is an alternative. The output data rate of each FESOC is 10 Gb/s or 20 Gb/s for 40-MHz or 80-MHz sampling rates, respectively. Each Front-End Board (FEB), which processes the signals of 128 detector cells, consists of 16 FESOCs and has 16 or 32 optical fibers operating at 10 Gb/s.

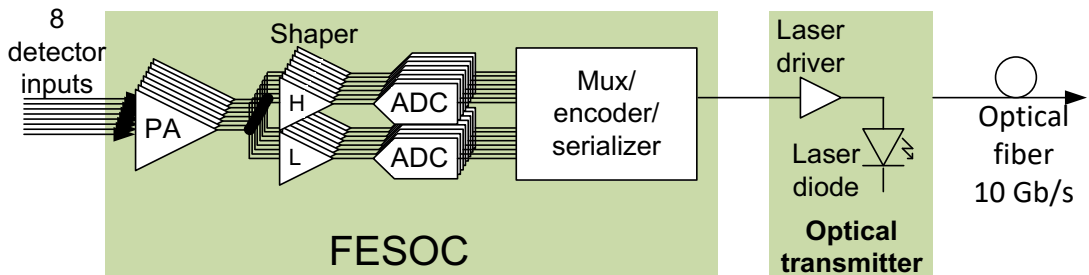

**Figure** 2. Block diagram of the proposed FESOC scheme

Several ASIC prototypes, including preamplifiers and shapers, ADCs, and laser diode array drivers, are under development for the front-end readout electronics upgrade. The designs and the simulation/test results of these prototypes will be described and discussed in this paper.

## 2. Development of preamplifiers and shapers

Three ASIC prototypes of preamplifiers and shapers are under development. The first two [8-9] are designed in 65-nm and 130-nm CMOS technologies. Both prototypes explore the possibility of implementing the low-power, low-noise preamplifiers and shapers in CMOS technologies. These two prototypes have similar design and test bench. Comparative measurements will be performed between these two prototypes. The third one is designed in a 180-nm SiGe BiCMOS technology.

All three prototypes must cover the input impedance, capacitance and current ranges specified as follows. The raw calorimeter signal is approximately a triangular current pulse with a rise time of 1 ns or less and a fall time of about 400 ns. The raw calorimeter signal is transmitted from the detector cell to a preamplifier via a 50-Ω or 25-Ω cable, depending on the detector capacitance. For the detector cells with 50-Ω cables, the maximum capacitance is 0.5 nF and the maximum raw current is 2 mA. For the detector cells with 25-Ω cables, the maximum capacitance is 2 nF and the maximum raw current is 10 mA.

All three prototypes also must meet the noise, Integral Non-Linearity (INL), and power consumption requirements. The Equivalent Noise Input (ENI) and the INL must be less than 60 nA and ±0.2% for the 2-mA range and 200 nA and ±0.5% for the 10-mA range, respectively. The power consumption must be less than 150 mW per channel. For comparison, the performances of the current preamplifiers and shapers include an ENI of 55 nA for a 400-pF detector capacitance and 150 nA for a 1.5-nF detector capacitance, an INL of ±0.2%, and a power consumption of 175 mW per channel.

### 2.1 Development in a 65-nm CMOS technology

The first prototype, HCL1, includes 8 channels. Each channel consists of a preamplifier and two shapers (high and low gains). HCL1 features programmable termination (±12% around 50 Ω or 25 Ω) and programmable peaking time (approximately from 30 ns to 50 ns). Figure 3 is the schematic of the preamplifier of HCL1. The core of the preamplifier is a fully differential amplifier. The ratio of the two capacitors determines the gain (G). In comparison to resistors, the use of noiseless capacitors reduces noise. The feedback resistor (R) and the gain (G) determine the input impedance, which is equal to R/(G+1), independent on the input current ($I_i$).

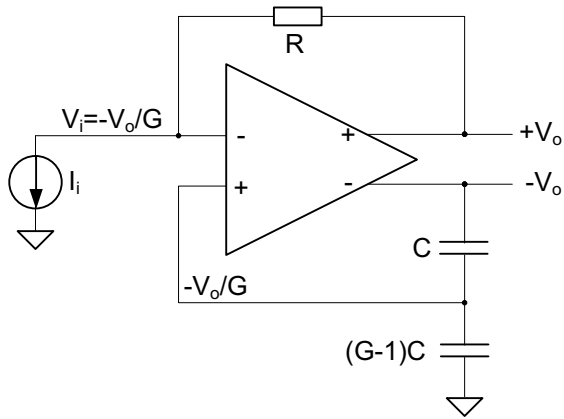
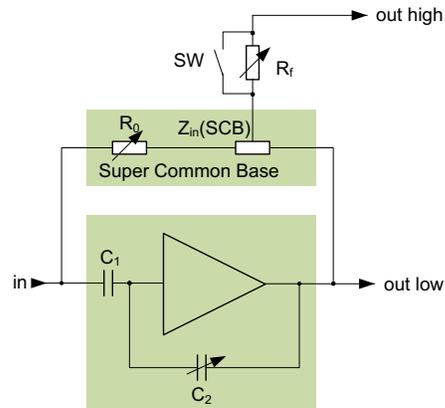

**Figure** 3. Schematic of the PA of HLC1    **Figure** 4. Schematic of the PA of LAUROC

The layout of HCL1 is being finalized and chip submission is imminent. All post-layout simulation results meet the design goal. The simulated ENI is 57 nA for a 260-pF detector capacitance and 160 nA for a 1.3-nF detector capacitance, respectively. The INL is within ±0.1% at the 0-9 mA range and within ±0.5% at the 0-10 mA range. HCL1 uses a single power voltage of 1.2 V with a power consumption of about 100 mW per channel. Each preamplifier consumes about 55 mW.

## 2.2 Development in a 130-nm CMOS technology

The second prototype, LAUROC, is designed in a 130-nm CMOS technology with the power voltage of 2.5 V. LAUROC has 8 channels of preamplifiers. The schematic of the preamplifier is shown in Figure 4. The preamplifier is a low-noise voltage-sensitive amplifier. To reduce the noise, the preamplifier uses noiseless capacitors to set the gain. The gain (G) is equal to $-C_1/C_2$, programmable by remotely turning on or off the switches connected to the capacitor $C_2$. The input impedance is equal to $(R_0+Z_{in}(SCB))/(1+|G|)$, determined by the resistor $R_0$, the input impedance of the Super Common Base (SCB) transistor, and the gain. The input impedance is programmable by tuning $R_0$ and the gain. The resistive noise, which is equal to $4kTR_0/(1+|G|)^2$ (k and T are Boltzmann Constant and the temperature, respectively), is much less than the normal thermal noise of $R_0$. In other words, the resistor is electrically cooled. The preamplifier has two outputs, one with high gain and the other one with low gain.

The layout of LAUROC is shown in Figure 5. The post-layout simulation with an ideal $CR-(RC)^2$ shaper shows that the ENI of the preamplifier is 46 nA for a 0.4-nF detector capacitance and 150 nA for a 1.4-nF detector capacitance, respectively. The INL is within ±0.2% in the input range of 10 µA to 10 mA. Each preamplifier consumes about 18 mW.

LAUROC has been fabricated and is packaged in a pin grid array (PGA) package (144 pins). The test is still ongoing. A picture of the test board is shown in Figure 6.

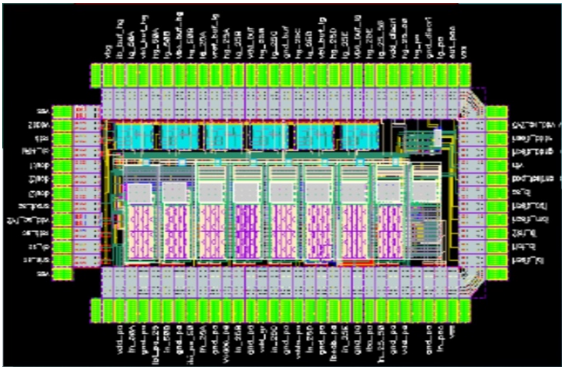 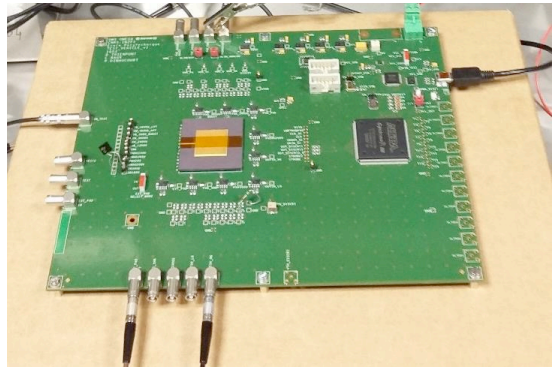

**Figure** 5. Layout of LAUROC    **Figure** 6. Test board of LAUROC

## 2.3 Development in a SiGe BiCMOS technology

The third prototype is designed in a 180-nm SiGe BiCMOS technology (IBM 7WL). Early design in SiGe BiCMOS technologies (IBM 8WL and IHP SG25H3P) achieved required INL and noise levels [10]. The prototype uses a similar design to the current one which is implemented with discrete components. The prototype has unipolar output pulse and two gain stages covering the desired dynamic range. The prototype provides bonding option for 25 or 50 Ω instead of impedance/dynamic range tuning. Simulation indicates that the design is marginal at high frequencies above 30 MHz. With $CR-(RC)^2$ shaping, the ENI is 32 nA for a capacitance of 0.2 nF and 97 nA for a capacitance of 1.0 nF, respectively. The design achieves an INL below ±0.1% in the input range of 0 to 9 mA. The power consumption is about 86 mW per channel.

Each preamplifier consumes 34 mW and 47 mW for the termination of 50 Ω and 25 Ω, respectively.

The layout of this prototype has been completed. However, the submission is still under discussion.

## 3. ADCs

The context study of the ADC ASIC has recently started. Figure 7 represents the block diagram of the proposed full ADC prototype. The prototype is designed in a 65-nm CMOS technology. The prototype includes 8 14-bit dynamic range ADC units, a current reference unit, a voltage reference unit, a Phase-Locked Loop (PLL) and clock unit which generates clocks from an input clock synchronous to the 40-MHz LHC clock, a clock distribution scheme, calibration engines, digital data processing units, a serializer system, a calibration/system control unit, and a slow control interface. The sampling rate is 40 MSamples/s. Each ADC unit includes a dynamic range enhancer (DRE) and a 12-bit Successive Approximation Register (SAR) ADC. The output digital data include 16 bits per channel per sample, implying that excess information such as Bunch-Crossing Identification (BCID) and a frame header could be encoded. Each chip produces a data volume of 5.12 Gb/s. In the prototype the digital data are serialized and transferred at 640 Mb/s (double data rate) to reduce the pin number.

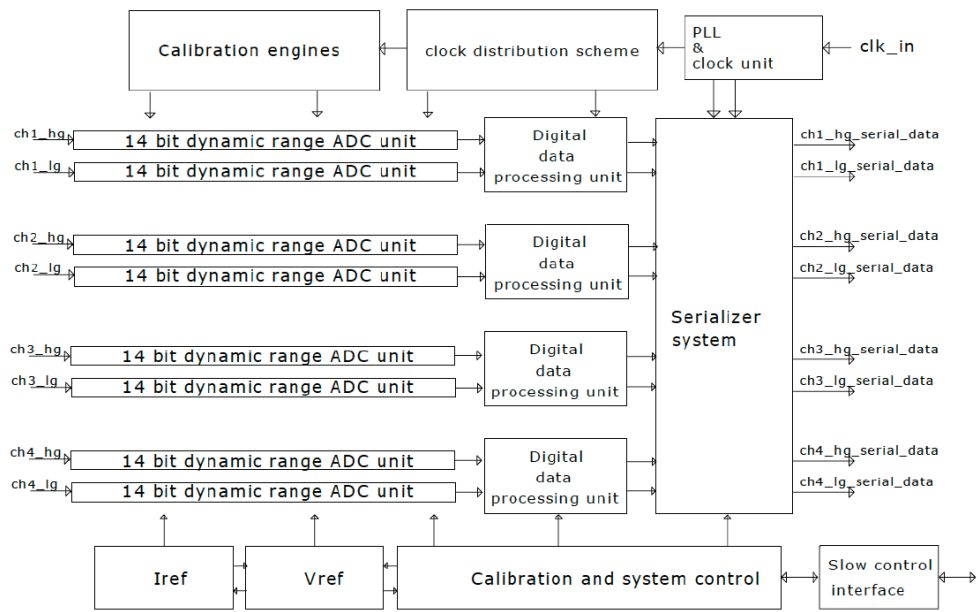

**Figure** 7. Block diagram of the proposed ADC prototype

The ADC development is in progress. The test chip including a DRE block and a SAR block will be submitted in April 2017.

## 4. Development of optical Links

The optical link part will utilize lpGBT [11] and Versatile Link + [12], aiming together at the development of radiation-tolerant optical links. The multiplexer, encoder and the serializer of lpGBT will be integrated with the preamplifiers, shapers and ADCs. In this section we will focus on the development of laser diode array drivers [13-14] for Versatile Link +.

Vertical-Cavity Surface-Emitting Laser (VCSEL) Array Driver (VLAD) and Low-Power VCSEL Array Driver (lpVLAD) [14] are quad-channel, 10-Gb/s-per-channel VCSEL array driver ASICs designed in a 65-nm CMOS technology with different output structures. Figure 8 is the schematics of VLAD and lpVLAD. The two amplifier stages on the left are pre-drivers that amplify the small input signal with a swing down to 100 mV (peak-peak) to a large signal with a swing larger than 600 mV. The pre-drivers are the same for VLAD and lpVLAD. The post-layout simulation indicates that bandwidth of the pre-drivers is larger than 12.5 GHz. The amplifiers on the right are the output drivers that drive the VCSEL with proper modulation current and bias current. For VLAD, the laser is turned off when the switching transistor $M_6$ is on and most current of the constant current source $I_3$ passes $M_6$. The laser is turned on when the switching transistor $M_6$ is off and most current of the constant current source $I_3$ passes through the laser diode $D_1$. For lpVLAD, the transistors $M_7$ and $M_8$ are turned on or off simultaneously and function as a transmission gate. The current passing through $M_7$ and $M_8$ provides modulation current for the laser diode. The constant current source $I_5$ provides the bias current for the laser diode $D_2$. As can be seen in the figure, the currents passing through $M_5$ and the current passing through $M_6$ are wasted in VLAD, resulting in a higher power consumption than lpVLAD.

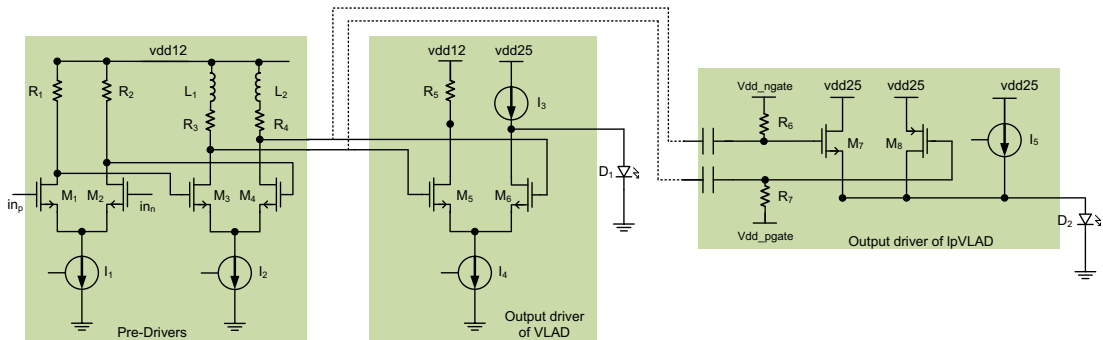

**Figure** 8. Functional block diagram of VLAD and lpVLAD

Figure 9 shows the layouts of VLAD and lpVLAD. Both dies are 1.9 mm x 1.7 mm. The pitch of the output channels is 0.25 mm, matching that of the VCSEL array.

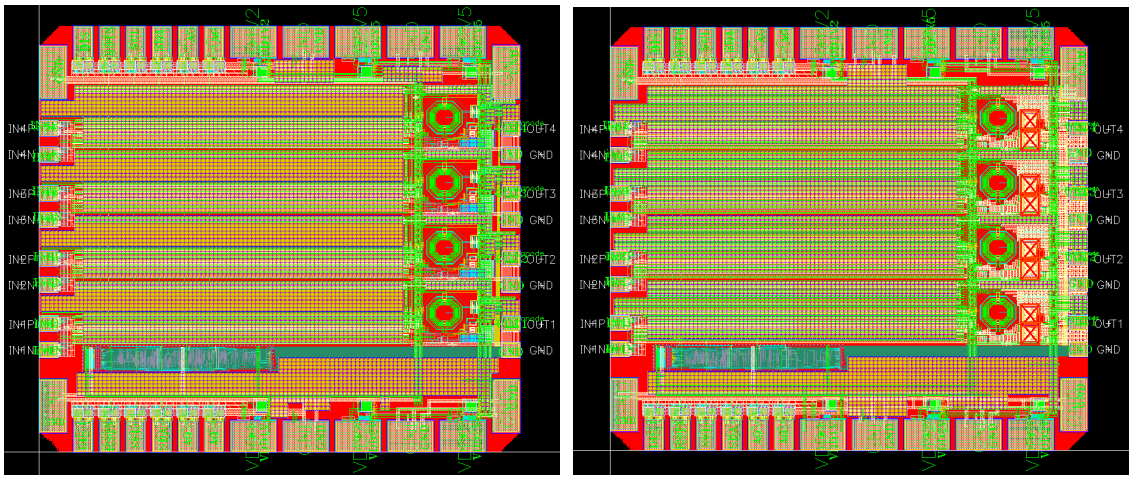

**Figure** 9. Layouts of VLAD (left) and lpVLAD (right)

Both VLAD and lpVLAD have been tested. During the test, a signal source (Model number J-BERT N4903B from Agilent Technologies) generated the input signals with a pattern of pseudo-random binary sequence $2^7$-1 and an amplitude of 400 mV (P-P) at 10 Gb/s. An oscilloscope (model number DSA91204A from Agilent Technologies) measured eye diagrams. An optical receiver (model number 81495 from Agilent technologies) converted the optical signals to electrical signals. The conversion efficiency was 3.25 mV/µW. The optical eye diagram of each channel was observed when two adjacent channels were working simultaneously.

The eye diagrams of VLAD and lpVLAD are shown in Figure 10. The bias current was 2.5 mA and the modulation current was 5.5 mA for VLAD and the bias current was 2.3 mA and the modulation current was 6.4 mA for lpVLAD. The measured total jitter was about 48 and 37 ps (Peak-Peak at the bit error rate of $1 \times 10^{-12}$) for VLAD and lpVLAD, respectively. The power consumption, including what the laser consumed, was 33.9 mW/channel and 22 mW/channel for VLAD and lpVLAD, respectively.

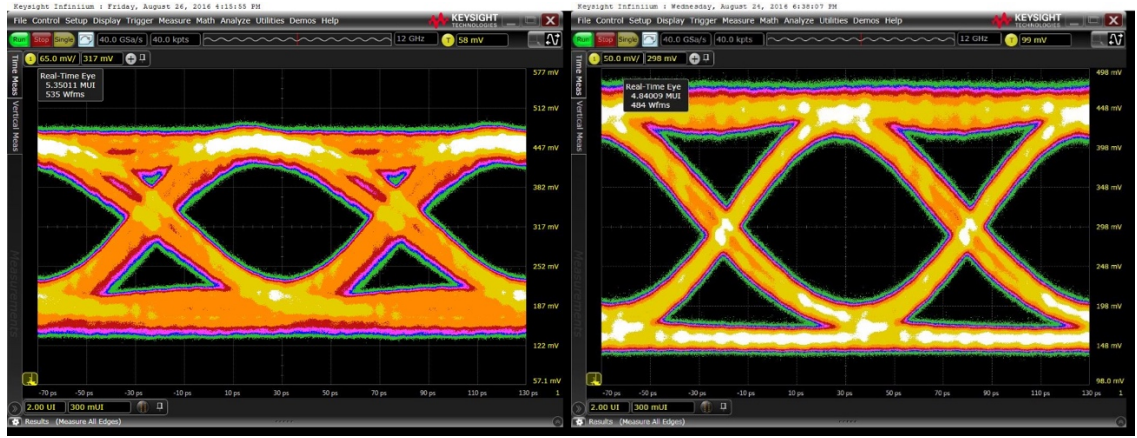

**Figure** 10. Optical eye diagrams of VLAD (left) and lpVLAD (right)

## 5. Conclusion

The ATLAS LAr front-end readout electronics without a trigger is under development to meet high luminosity requirements. An approach of FESOC, in which preamplifiers, shapers, ADCs, multiplexers, encoders and serializers are integrated on a single silicon die or in a single package, is being pursued. Three preamplifier/shaper ASICs in early development stages show promising performances within termination, capacitance, noise, INL, dynamic range, and power consumption requirements. ADC design has started. Furthermore, results of two newly designed VCSEL array driver ASICs show that the required 10Gb/s data rate at 22-34 mW per channel is achieved, suitable for integration into a low-power optical link package.